\documentstyle[aps,12pt]{revtex}

\newcommand{\DATE}  {\today}

\newcommand{\PPrtNo}
{
MSU-HEP-60508 \\ CTEQ-605
}
\newcommand{\TITLE}
{
Comparison of CDF and D0 Inclusive Jet Cross-sections 
}
\newcommand{\THANKS}
{
This work was partially supported by NSF. \\
~~This short CTEQ note contains material extracted from a talk given at the Rome DIS conference in April, 1996. It is presented here because of the strong current interest in the implications of the CDF and D0 inclusive jet data. Some of these results will appear in our contribution to the Proceedings of the Rome DIS conference, and in a forthcoming paper on a systematic global analysis including new DIS and jet data resulting in a new series of CTEQ4 parton distributions \cite{jet2}.
}
\newcommand{\AUTHORS}
{
H.~L.~Lai and W.~K.~Tung
}
\newcommand{\INST}
{
Department of Physics and Astronomy, Michigan State Univ., E. Lansing, MI
}
\newcommand{\ABSTRACT}
{
\begin{abstract}
The recently reported CDF and D0 inclusive jet cross-sections are compared, using a uniform theoretical NLO QCD calculation to account for the different kinematic coverages of the pseudo-rapidity variable in the two experiments. The two data sets are found to be in good agreement. With a 2-3\% relative overall normalization adjustment, the data sets appear to agree over the entire $E_t$ range, even without taking into account the other systematic errors. 
\end{abstract}
}

\begin{document}

\begin{titlepage}

\begin{tabular}{l}
\DATE
\end{tabular}
\hfill
\begin{tabular}{l}
\PPrtNo
\end{tabular}

\vspace{2cm}

\begin{center}
\renewcommand{\thefootnote}{\fnsymbol{footnote}}
{
\LARGE \TITLE  \footnote[2]{\THANKS}
}
\renewcommand{\thefootnote}{\arabic{footnote}}

\vspace{1.25cm}
{\large  \AUTHORS}

\vspace{1.25cm}

\INST
\end{center}

\vfill

\ABSTRACT                 

\vfill

\newpage
\end{titlepage}


A great deal of attention has been given recently to the high statistics
inclusive jet production measurements made at the Tevatron, stimulated by
the observation by the CDF
collaboration of a larger cross-section at high jet $E_t$ than expected from NLO QCD calculations based on previously
available parton distributions. \cite{cdfIa,cdfIb} This result may have far
reaching consequences if it is confirmed experimentally, and if it cannot be
explained in the conventional theoretical framework. Thus, the recently
reported independent measurement by the D0 collaboration \cite{d0Iab} was
anticipated with a lot of interest. Unfortunately, the comparison of the
results from the two experiments has, so far, led to rather ambiguous
interpretations. This is partly due to the fact that, although the
statistics are high for most of the measured $E_t$ range, the systematic
errors on these preliminary data are too large to allow for a definitive
conclusion. These systematic errors have yet to be fully analyzed and
properly taken into account in a meaningful comparison. If one overlooks the
systematic errors, the comparison plots displayed \cite{d0Iab} leaves the
impression that the two sets of data disagree in general shape, as well as in
normalization over the well-measured medium $E_t$ range of $100-200$ GeV.

There is a second source of uncertainty in the comparison: the two experiments
have slightly different kinematical coverage in the pseudo-rapidity
variable---$0.1<|\eta |<0.7$ for CDF vs. $|\eta |<0.5$ for D0. This makes it
impossible to compare the measured cross-sections directly because the
cross-section has an $\eta $ dependence in general. This difference in $\eta 
$ coverage must be corrected before a meaningful comparison can be made. The
correction factor can only be generated from some theory. Since NLO QCD is
very successful in accounting for the measured cross-section over 8 orders
of magnitude in the observed $E_t$ range, it is then natural to use the NLO
QCD theory as the common meeting ground for comparing the two experiments.
In practice, one computes the percentage difference between the individual
measurement versus the respective NLO QCD theory expectation, and then compares
the two differences. Thus, effectively, one is comparing the two
experimental results, normalized to theory. However, in the comparison of
the two data sets presented previously, the theoretical corrections were
calculated separately by the two collaborations using two different NLO QCD
programs: EKS \cite{EKS} for CDF and JETRAD \cite{JetRad} for D0. Since the
application of these NLO programs is known to be a delicate matter
(involving jet algorithms, scale choices, jet merging prescriptions, ...
etc.), it cannot be taken for granted that the existing comparisons are
unambiguous, even if the two programs have been checked against each other
under other circumstances before. The possibility exists that, in comparing
(Data - ``Theory'') / ``Theory'' from the two experiments, the ``Theory''
were not the same---hence the comparison was not an appropriate one.

Thus, we have undertaken to do an independent comparison of the two measured
cross-sections using an {\em uniform theoretical
calculation} as the common calibration. Specifically, we used a
recent version of the EKS program to calculate the expected ``Theory''
values for both the CDF and D0 data points, integrating over their
respective $\eta $ ranges, and then compared the two (Data - Theory) /
Theory results.\footnote{The renormalization and factorization scales 
are both set to $\mu = E_t/2$.
The theoretical cross-section is quite insensitive to the choice of $\mu$
for the range of $E_t$ considered. \cite{EKS} \cite{JetRad} }
We found, surprisingly, that {\em the two sets of data
agree rather well with each other}. Although it is not possible to quantify
the agreement/disagreement (say, by a meaningful $\chi ^2$), without a
proper treatment of all the correlated systematic errors, 
the impression gained from these results is qualitatively
different from that mentioned before, as we will now show.

Fig.~1 and Fig.~2 present the comparison of preliminary CDF and D0 Run-IB data
normalized to theory, using the CTEQ3M \cite{cteq3m} and MRSD0' 
\cite{mrsd0} parton distribution sets,
respectively, which were obtained before the jet data became available. The
error bars are statistical only. In these figures, one sees the, by now,
well-known higher-than-expected CDF cross-section. However, the two sets of
data seem to be in rather good qualitative agreement. 
In the medium $E_t$ range (say,
$50-200$ GeV) where the cross-section is well-measured, there appears to be a
2-3\% difference in relative overall normalization between the two experiments,
which is well within the experimental normalization uncertainties, but quite a
bit smaller than that seen in earlier comparisons. If an
adjustment of the relative normalization of this magnitude is made, the
agreement will look even better (see next paragraph). It may be tempting to notice a slight difference in slopes of the two data sets as a function of $E_t$. But one must bear in mind that correlated errors are not included in this comparison. Several of these systematic errors can easily lead to $E_t$ dependent corrections which will nullify the observed effect.
 
Similar conclusions are reached using as reference two recent parton
distribution sets which incorporate some of the jet data in the global fit.
Fig.~3 uses the forthcoming CTEQ4M distributions \cite{jet2} which includes
the medium range $E_t$ jet data, along with the most recent H1 \cite{H196} and
ZEUS \cite{Zeus96} deep inelastic scattering measurements; whereas Fig.~4 uses
one of the CTEQHJ parton sets which are tailored to accommodate the CDF high
$E_t$ ($>200$ GeV) jets along with the other data sets \cite{jet1}. In this
comparison, an overall normalization shift between data and theory was allowed
during the fit for each experiment.  The resulting normalization factors for
CDF/D0 were found to be 1.01/0.99 for CTEQ4M and 1.01/0.98 for CTEQHJ.  These
normalization factors were applied to the data points in these figures. Thus, there is a relative normalization shift between the two sets of
data by about 2-3\% for both of these plots. We see that these new parton
distributions give better fits to the jet data; and they provide the same
conclusions concerning the comparison of the CDF and D0 data relative to each
other. Fig.~4 particularly highlights the remarkable agreement between the two experiments over the entire $E_t$ range, even when all systematic errors (except overall normalization) have been left out.

These simple calculations show that, on one hand, it is gratifying to see the agreement between the inclusive jet cross-sections measured by CDF and D0
when a uniform
theoretical calculation is used to correct the different $\eta $ coverages.
On the other hand, the fact that our results differ from previous
comparisons underlines the sensitivity of the NLO QCD calculation of jet
cross-sections to subtle effects of jet algorithms, scale-choice, and
delicate cancellations among various contributions, which all have to be
handled with care if precision at a few percent level is required. It is possible that the same program can
give different answers with different parameter choices; and different versions
of the same program may not give the same answers if not suitably adjusted.

Since the results described here were presented at the Rome DIS conference,
a number of concerted efforts are being made by the various groups to study the sensitivity of the theoretical calculation to the various factors mentioned above, to the accuracy necessary for a full understanding of all the data and their physics implications. 

\vspace{1.0cm}
\newpage
\noindent {\bf Acknowledgment}: We thank Joey Huston, Steve Kuhlmann, D. Soper and Harry Weerts for many helpful discussions, comments and suggestions; 
D. Soper for providing the EKS program and advice on its use; 
and Anwar Bhatti and Jerry Blazey for discussions and for providing the experimental cross-sections.


\input epsf
\newpage

\begin{figure}[tbh]
\epsfxsize=6in
\epsfbox{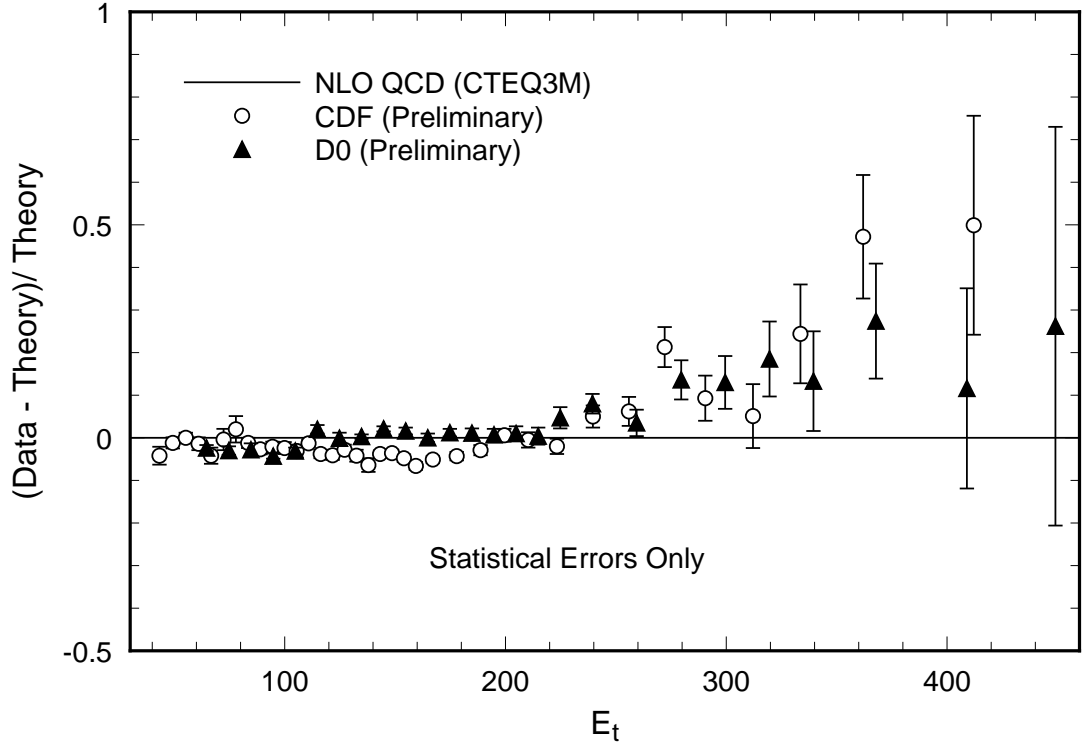}
	\caption{The preliminary CDF and D0 Run Ib data compared to NLO QCD using CTEQ3M parton distributions.}
	\label{d0cdf-3m}
\end{figure}
\begin{figure}[tbh]
\epsfxsize=6in
\epsfbox{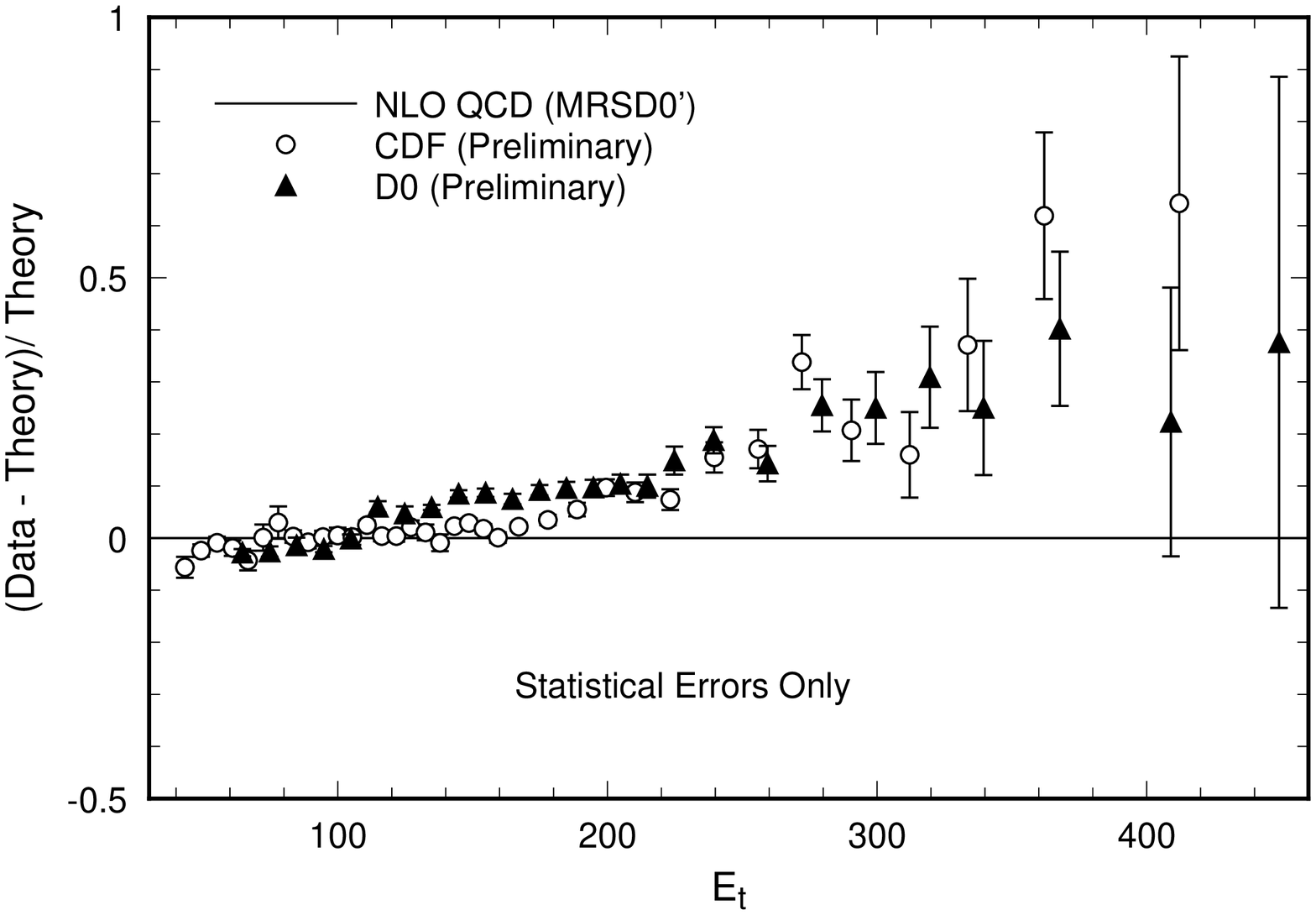}
	\caption{The preliminary CDF and D0 Run Ib data compared to NLO QCD using MRSD0' parton distributions.}
	\label{d0cdf-mrsd0}
\end{figure}
\begin{figure}[tbh]
\epsfxsize=6in
\epsfbox{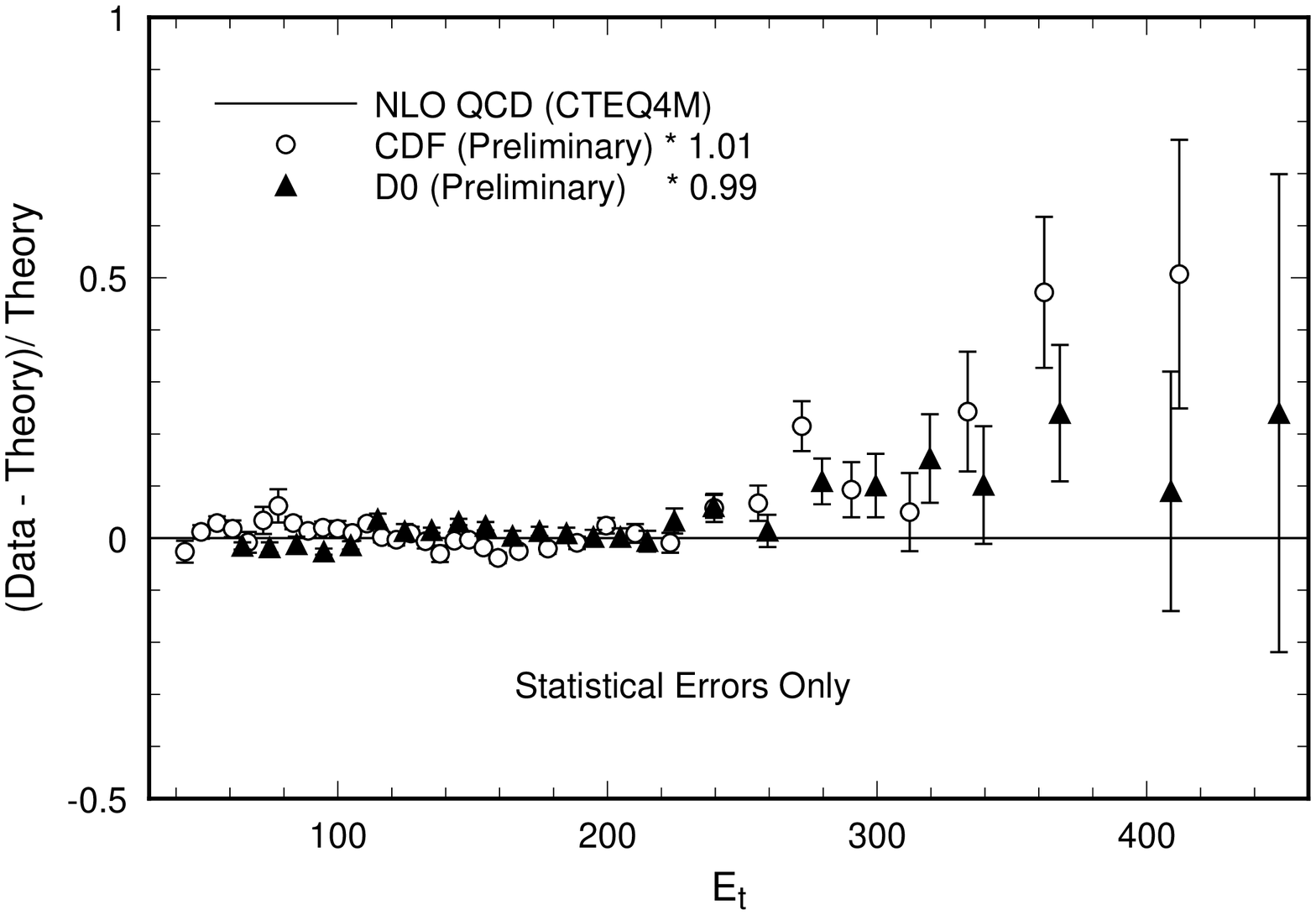}
	\caption{The preliminary CDF and D0 Run Ib data compared to NLO QCD using CTEQ4M parton distributions. Experimental points normalized as indicated.}
	\label{d0cdf-4m}
\end{figure}
\begin{figure}[tbh]
\epsfxsize=6in
\epsfbox{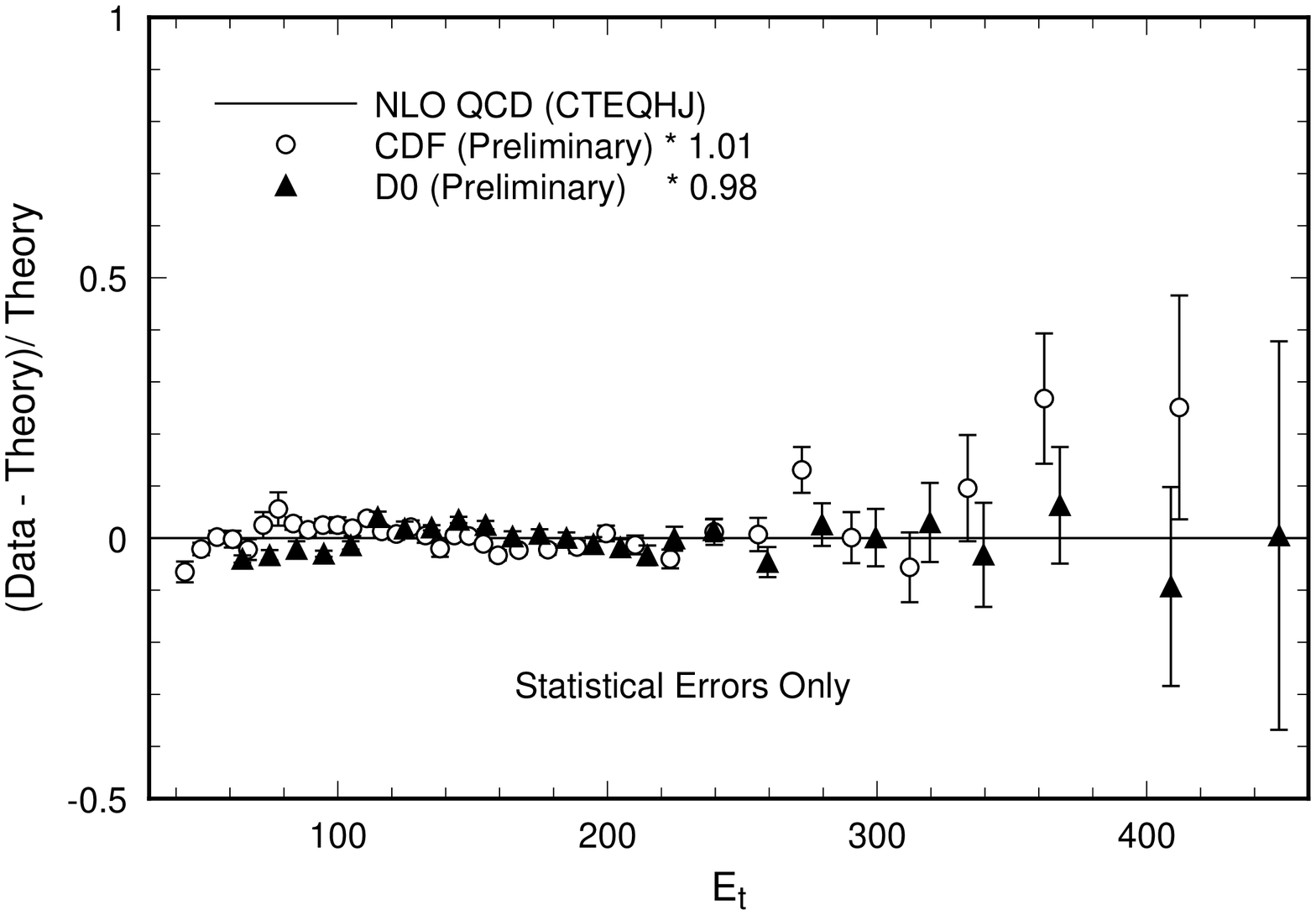}
	\caption{The preliminary CDF and D0 Run Ib data compared to NLO QCD using CTEQHJ parton distributions. Experimental points normalized as indicated.}
	\label{d0cdf-hj2}
\end{figure}

\end{document}